\documentclass[12pt]{article}

 \usepackage{amsfonts}
 \usepackage{amssymb,amsmath}
 \usepackage{graphicx} \usepackage{epstopdf}
 \newcommand{\crla}[1]{\quad\hbox{\scriptsize{#1}}\label{#1}\\[2pt]}
 
 \newcommand{\crlc}[1]{\quad\hbox{\scriptsize{#1}}\label{#1}}
 
 \newcommand{\eela}[1]{\quad\hbox{\scriptsize{#1}}\label{#1}\end{eqnarray}}
 \newcommand{\eelb}[1]{\label{#1}\end{eqnarray}}
 \newcommand{\newseca}[2]{\section{#1\ {\scriptsize #2}}\label{#2}\setcounter{equation}{0}}
 
 \newcommand{\bibitema}[1]{\bibitem{#1}{\scriptsize #1}}
 \newcommand{\citea}[1]{\cite{#1}{\scriptsize #1}}
 \newcommand{\seelabels}{\def\eel{\eela}\def\eeql{\eeqla} \def\crl{\crla} \def\newsecl{\newseca}\def\bibiteml{\bibitema}\def
 \citel{\citea}\def\labell{\crlc}}
 
\newcommand{\eeqla}[1]{\quad\hbox{\scriptsize{#1}}\label{#1}\end{aligned}\end{equation}}
\newcommand{\eeqlb}[1]{\label{#1}\end{aligned}\end{equation}}

\newcommand\testprintversion{\seelabels\setlength{\textheight}{9.3in}\setlength{\oddsidemargin}{0in}
   	 \setlength{\textwidth}{6.3in}\setlength{\topmargin}{-1in}}

\def\beq{\begin{equation}\begin{aligned}}		\def\eeq{\end{aligned}\end{equation}}
\def\be{\begin{eqnarray}}  					\def\ee{\end{eqnarray}}		
\def\bi#1{\begin{itemize}\item[#1]} 			 			\def\ei{\end{itemize}}
  \def\eqn#1{(\ref{#1})}
   	 \def\fn{\footnote}	  		

		       \def\del{\delta}        
		 \def\Lam{\Lambda}            

	    		        		     		\def\vv{\varphi}     
 	 		     	\def\tht{\theta}      	
	     	         
\def\W{\Omega}    		  		\def\dd{{\rm d}} 		
\def\OO{{\mathcal O}}

\def\ffract#1#2{\raise .2 em\hbox{$\scriptstyle#1$}\kern-.3em/\kern-.2em\lower .15 em \hbox{$\scriptstyle#2$}}

\def\bpmatrix{\begin{pmatrix}} 			\def\epmatrix{\end{pmatrix}}
\def\bmatrix{\begin{matrix}} 			\def\ematrix{\end{matrix}}
\def\bcenter{\begin{center}}			\def\ecenter{\end{center}}
  
\def\inn{{\mathrm{in}}}  \def\outt{{\mathrm{out}}} 
      \def\BH{{\mathrm{BH}}}     

\def\lowerheightfig#1#2#3{\(\raise-#1\hbox{\includegraphics[height=#2]{#3}}\)}
\def\lowerwidthfig#1#2#3{\(\raise-#1\hbox{\includegraphics[width=#2]{#3}}\)}
		 
\def\weglaten#1{ }

				\testprintversion

\begin{document}

\begin{titlepage}
 \title{\vskip 20mm \LARGE\bf Black Hole Firewalls and Quantum Mechanics}
\author{Gerard 't~Hooft}
\date{\normalsize Institute for Theoretical Physics \\ Utrecht University  \\[10pt]
Princetonplein 5 \\
3584 CC Utrecht \\
 the Netherlands  \\[10pt] internet:
http://www.staff.science.uu.nl/\~{}hooft101/}
 \maketitle

\begin{quotation} \noindent \text\bf {Abstract } \medskip \\	
Firewalls in black holes are easiest to understand by imposing time reversal invariance, together with a unitary evolution law.  The best approach seems to be to split up the time span of a black hole into short periods, during which no firewalls can be detected by any observer. Then, gluing together subsequent time periods, firewalls seem to appear, but they can always be transformed away. At all times we need a Hilbert space of a finite dimension, as long as particles far separated from the black hole are ignored. Our conclusion contradicts other findings, particularly a recent paper by Strauss and Whiting.  Indeed, the firewall transformation removes the entanglement between very early and very late in- and out-particles, in a far-from-trivial way.
\end{quotation} \vfill
 Version 2, \today 

 \end{titlepage} 
\setcounter{page}{2}

One crucial step towards understanding quantum gravity, is to find a logically accurate description of quantum particles near the horizon of a black hole. The problem seems to be easier than formulating the entire theory, whereas it is sufficiently non-trivial to teach us important things. Thus, the black hole problem is of great importance, and we should proceed with utmost care, step-by-step, with our observations and calculations.

In several papers\cite{GtHfirew.ref, GtHErice.ref}, the present author proposed\fn{In Ref. \cite{GtHErice.ref}. regions \(I\) and  \(II\) of the Penrose diagram for a stationary black hole were also proposed for describing the \emph{antipodes} of one another. We now have doubts about this, as it appears to lead to conflicts in the sign switches for even and odd \(\ell\) states\cite{GtHinsideout.ref}.}  to use a general procedure for describing the Hilbert space of a quantum black hole.  We found that particles moving into a black hole (``in-particles") generate curvature at the past event horizon, after a time lapse larger than
\(\OO(GM\log M)\) in natural units. For classical particles, this would be merely  a coordinate redefinition, but in the case of a quantum in-particle, the situation is more subtle.

For the out-particles, using time reversal symmetry, we can say the same thing. We called the singularities on the horizons thus obtained ``firewalls", since in our view these are exactly the firewalls discussed
in the literature\cite{firewalls.ref,Susskind.ref}. We named them the same way, although most authors are more concerned about the quantum entanglement arising at much longer time scales.\cite{Page.ref}

This author's point of view is that a mechanism has been found  that removes the firewall (both at short and at very long time scales), and this is essential to provide more information for the construction of a theory uniting gravity with quantum mechanics.
The procedure used is that in Hawking's original arguments, Hawking particles do not generate any firewall. Thus no firewalls should be seen to be produced by out-particles. Time reversal symmetry should then allow us to say the same for in-particles. Indeed it does, but we have to understand what we are doing. It is highly non-trivial.

What we call  `firewalls'  can be moved from future event horizon  to past event horizon and back. This, we claim, is a transformation that restores the symmetry of the dynamical rules under time reversal,  where most investigators encounter a singularity that they cannot quite handle, as it seems. This author's point of view is that finding a mechanism  that removes the firewall will be essential to provide more information for the construction of a theory uniting gravity with quantum mechanics, while safeguarding unitarity. We claim that this also removes the entanglement problem, which appears to have been standing in the way of further progress\cite{firewalls.ref, Susskind.ref,Page.ref}.

Our procedure handles the firewall naturally. It works beautifully, but, as we shall explain, it also shows how hard quantum gravity will be -- the entire problem has of course not been solved.

	In a recent paper, Strauss and Whiting\cite{StraussWhiting.ref} dismiss the transformation used by us, as it seems to be `classical'  to them, and they arrive at a `more accurate' procedure where firewalls do continue to cause havoc. Here, we reiterate how our firewall transformation works, and why it is to be regarded as a purely quantum feature. Having a difference of opinion simply reflects the fact that a complete and universal treatment of quantum gravity is still missing, but we shall illuminate what can be done with firewall transformations, and how they completely remove the singularities at the  future and past horizons. Instead of a singularity, what we end up with  is a boundary condition that bounces all in going particles  outwards, as explained below.
	
Since we are interested in understanding what happens very close to the crossing point of the two horizons, we may assume the particles to move in and out radially, since the gravitational effects of motion in the transverse directions are subdominant.

Also, if the black hole is large and heavy in Planck units, the momenta and energies of the particles considered may be assumed to be much smaller than the black hole mass and energy. This is why we only keep the dominant radial gravitational effects that in-particles have on out-particles and \emph{vice versa}.
	
We then consider the task of setting up the black hole Hilbert space. If we ignore particles far separated from the black hole, this Hilbert space should become finite-dimensional (the number of its orthonormal basis elements is the exponential of the entropy, which is known to be finite).  We find that it is not only finite, it is very small. If we estimate which particles in the surrounding Unruh state\cite{Unruh.ref}  may contribute to the entropy, we find that it is only those that are separated from the horizon by a distance comparable to or larger than the Planck length.\cite{GtHquantstruct.ref}  This important observation will be used below to remove the firewalls.

We now make use of a central observation. Particles moving into a black hole (the `in-particles'), at a solid angle \(\W=(\tht,\vv)\),  carry a momentum distribution \(p^\inn(\W)\) that will have a gravitational effect on the `out-particles', such that the spacetime point \(u^\outt(\W)\) where the out-particles cross the past horizon, is shifted:
\be	\del u^\outt(\W)&=&\int\dd^2\W'f(\W,\W')\,p^\inn(\W')\ , \labell{outin.eq}\\
	\del u^\inn(\W)&=&-\int\dd^2\W'f(\W,\W')\,p^\outt(\W')\ , \eel{Shapiro.eq}
	
These equations are called after Shapiro\cite{Shapiro.ref} just because the calculation uses the same math as that used by Shapiro to calculate this effect. Curiously, Strauss and Whiting\cite{StraussWhiting.ref} complain that these equations are ``only classical", while they are completely linear, and they invite the researcher to impose the usual commutation rules between momenta \(p\) and positions \(u\). Clearly, these equations allow for superpositions. We have genuine quantum mechanics here (some further remarks on this are made near the end of this paper).

The fact that the commutator \([u^\inn(\W),\,u^\outt(\W')]\) changes sign upon interchanging
\(\hbox{in\(\leftrightarrow\)out}\), explains the need for the minus sign in Eq.~\eqn{Shapiro.eq}.

 It is important to note\fn{Perhaps this was not understood in Ref.~\cite{StraussWhiting.ref}. Remarkably, this redefinition does not affect the quantum commutators \eqn{commrel.eq}. Thus, our treatment is definitely quantum mechanical.}
	 that \(p^{\inn,\,\outt}(\W)\) refer to  \emph{total momentum density} at the transverse point 
	 \(\W=(\tht,\vv)\) and not just a single particle, and similarly, \(u^{\inn,\,\outt}(\W)\) refers to the \emph{average radial position} of the particles at the point \(\W\).
 
In Eqs.~\eqn{outin.eq},\,\eqn{Shapiro.eq}, the symbol \(\del\) indicates that the \emph{change} of the position of the out-particles is caused by the \emph{total} momentum of the in-particles and vice versa. This subtlety might have been missed by Strauss and Whiting.  Indeed, we subsequently erase the delta \((\del)\). The reason for this is a fundamental one: 
\emph{the earliest out-particles are moved much more strongly than the later ones,} following exactly the fact that the \emph{momenta} of the early \emph{in-particles} have a much bigger effect than that of the later ones. 
Therefore, a cut-off is needed: the earliest particles to be considered are the earliest ones allowed in our Hilbert space. After all, our Hilbert space at any moment of time only describes particles visible at that time period. Anything going in or out much earlier (or much later) than that is assumed to have \(p^\inn=0\) and \(u^\outt=0\). So from that point on we identify \(u^\outt\) as describing positions of all out-particles \emph{that we are allowed to consider.}

Our disagreement with Ref.  \cite{StraussWhiting.ref}, is presumably due to the fact that they overlooked this essential step in our procedure. It means that the `firewall transformation' indeed affects Hilbert space in a non-trivial way: the earliest in-particles have become invisible, but they left their footprints in the out-particles, which implies that the information they carried has not disappeared.
Omitting these early in-particles in our Hilbert space is just prescribed to avoid over-counting. Thus we see that our firewall transformation non-trivially affects the dynamics.  The momenta of the latest out-particles appear to form an impenetrable ingredient of a firewall that puzzled earlier authors. Our proposal is that, indeed, we must remove this firewall. For that, we now use the equations \eqn{outin.eq} and \eqn{Shapiro.eq} to \emph{replace} the out-particles by the in-particles and back, or equivalently, \emph{identify} in- and out-particles. 

There is nothing wrong with this. What we get is that, as the earliest out-particles are \emph{generated} by the earliest in-particles, we see that large \(p^\inn\) is \emph{replaced} by the earliest \(u^\outt\) and now, we simply conclude: yes, \(u^\outt\) is very large. Of course it is, the earliest out-particles already left the black hole long ago, so that their position operators \(u^\outt\) are large. Removing the firewall caused by the out-particles, is tantamount to ignoring the out-particles that now are too far away to cause any harm. 

Now that we arrived at this point, we can frame the situation more accurately. As we removed
 the \(\del\), we now have a simple \emph{identification} connecting the in-momenta to the out-positions and \emph{vice versa}. We subsequently ignore all particles that have coordinates \(u\) 
that are outside the domain \([-\Lam,\,\Lam]\), where \(\Lam\) is a reasonably chosen boundary, simply indicating that we wish to ignore particles that are too fas separated from the black hole. Of course this modifies our Hilbert space, but in a way anybody should understand. Since the coordinates \(u\) are now limited to this domain, Eqs.~\eqn{outin.eq} and \eqn{Shapiro.eq} also command that the momenta \(p\) for both the in- and the out- particles are limited to a similar domain. As is well known, restricting a variable as well as its Fourier transform to a finite domain, implies that the dimensionality of the entire Hilbert space now is finite! \emph{That} removes the entanglement problem: the out-particles carry the information of the in-particles just as in any quantum scattering problem. Therefore we can regard the finite Hilbert space of the black hole with those particles that are still very close, just as the various quantum numbers of the excited states and bound states of the black hole.

As time proceeds,  new in-particles move into their allowed domain and out-particles leave the scene and are crossed out. The black hole contribution to Hilbert space keeps its finite size, but in-particles too close to the horizon, automatically change into out-particles, by Eq.~\eqn{outin.eq}.

Thus, by \emph{identifying} \(u^\outt\) by \(p^\inn\) we removed the firewall. Our replacement procedure acts as a \emph{boundary condition} at the origin where the horizons meet: in-particle becomes out-particle. The domain where this boundary condition acts, has the dimensions of a Planck length squared.

One very important question must now be posed: \emph{why do we have two outside regions,} region \(I\) where all \(u>0\) and region \(II\) where all \(u<0\) \ ?

There are several ways to handle this question. One proposal was to identify the transformation
 \(u\leftrightarrow -u\) to the \emph{antipodal} transformation. This however generates a minus sign for all odd values of the partial wave quantum \(\ell\) that causes problems that we rather avoid. We find the following resolution of this problem more elegant:\\
 \indent\emph{Region \(II\) is an exact quantum clone of region \(I\).}\cite{GtHclones.ref} This becomes natural if we consider a replacement of the Kruskal-Szekeres coordinate frame by the Schwarzschild coordinates. In Schwarzschild, the regions \(I\) and \(II\) exactly coincide.

However, there may be other misunderstandings. We claim that our description is entirely quantum mechanical. Indeed, the positions and the momenta must obey the usual commutation relations 
\be [u(\W),\,p(\W')]=i\del^2_{\W,\W'}\ . \eel{commrel.eq} 
This however appears to turn these variables back into operators acting on single particles, rather than quantised fields. It is a puzzling feature why this is so, but we claim it is totally inevitable. Gravity \emph{only} acts on momenta and only affects positions (by curving space and time). Attempts to replace \(u(\W)\) and \(p(\W)\) by quantised fields do not work at the horizon crossing point. This implies that a delicate transformation will be needed to turn the particles far away from the black hole, into fully licenced Standard Model 
particles. This is a very important puzzle. We do not think there is any contradiction here, the problem is just a hard one, and has not yet been fully addressed.

A consequence of the points raised above, is that we can improve upon the approximation made by Strauss and Whiting. Why did they limit themselves to spherically symmetric mass shells? We found that the equations \eqn{outin.eq} and \eqn{Shapiro.eq} are ideally suited to a spherical wave expansion, replacing \(u(\W)\) and \(p(\W)\) by \(u_{\ell,m}\) and \(p_{\ell,m}\). The different \(\ell,m\) components decouple entirely, turning our dynamics into one-dimensional quantum mechanics, at each \((\ell,m)\) pair, which is easy to solve.\cite{GtHfirew.ref}--\cite{GtHinsideout.ref} This we find to be a very powerful positive feature of our proposal.

We emphasise that the way Hilbert space basis elements  are counted in our procedure, applies to all sequences of events within a time interval of the order of \(M_\BH\log{M_\BH}\) in natural units. For describing different periods of time one merely has to apply the relevant firewall transformations in such a way that the particles that are relevant in the time lapse considered are all made visible. The particles that would form firewalls will be invisible so that the firewall disappears from our descriptions.

Also the \emph{Page time}\cite{Page.ref}, often used to keep track of entangled particles, does not appear to play a role in the treatment here, as we insist  to only consider much shorter time intervals.

Any questions concerning \emph{entanglement} and \emph{information exchange} between  early and late larticles, have the same answers as when our boundary condition near the horizons were replaced by a \emph{brick wall} close to the horizon, and the particles would be just as an ideal gas of spin zero particles without chemical potentials. The position of the brick wall can be adjusted to match with an atmosphere at the right temperature\cite{GtHquantstruct.ref}.

Firewalls appear to play a role in many investigations\cite{Susskind.ref}, and yet this author believes that they should be taken care of the way described here.

We thank N. Gaddam and U. Gursoy for useful discussions and remarks. Also N. Strauss and B. Whiting for writing up their considerations and findings, even if we do not agree. It is the kind of discussions that may lead us into considering the next steps.

\end{document}